\documentclass[english]{aa}

\usepackage{natbib}
\usepackage{amssymb}
\usepackage{graphicx}

\bibpunct{(}{)}{;}{a}{}{,} 

\newcommand{\be}{\begin{equation}}
\newcommand{\ee}{\end{equation}}
\newcommand{\ba}{\begin{eqnarray}}
\newcommand{\ea}{\end{eqnarray}}

\newcounter{IonCS}
\renewcommand{\ion}[2]{\setcounter{IonCS}{#2}#1\,{\small{\Roman{IonCS}}}}
\newcounter{RomC}
\newcommand{\rom}[1]{\setcounter{RomC}{#1}{\Roman{RomC}}}

\begin{document}

\title{Dynamics of solar coronal loops}
\subtitle{\rom{2}. Catastrophic cooling and high-speed downflows} 

\author{D.A.N. M\"uller\inst{1,2},  H. Peter\inst{1} \and V. H. Hansteen\inst{2}}

\titlerunning{Dynamics of solar coronal loops {\rom{2}}}
\authorrunning{M\"uller et al.} \offprints{D.A.N.\ M\"uller}

\institute{Kiepenheuer-Institut f\"ur Sonnenphysik, Sch\"oneckstr. 6, D-79104 Freiburg, Germany\\ \email{Daniel.Mueller@astro.uio.no, peter@kis.uni-freiburg.de} \and Institute of Theoretical Astrophysics, University of Oslo, P.O. Box 1029, Blindern N-0315, Oslo, Norway\\
\email{Viggo.Hansteen@astro.uio.no} }

\date{Received 8 March 2004; accepted 14 May 2004}

\abstract{
This work addresses the problem of plasma condensation and ``catastrophic cooling'' in solar coronal loops. We have carried out numerical calculations of coronal loops and find several classes of time-dependent solutions (static, periodic, irregular), depending on the spatial distribution of a temporally constant energy deposition in the loop. Dynamic loops exhibit recurrent plasma condensations, accompanied by high-speed downflows and transient brightenings of transition region lines, in good agreement with features observed with TRACE.
Furthermore, these results also offer an explanation for the recent EIT observations of \cite{DeGroof+al2004AA} of moving bright blobs in large coronal loops.
In contrast to earlier models, we suggest that the process of catastrophic cooling is not initiated by a drastic decrease of the total loop heating but rather results from a loss of equilibrium at the loop apex as a natural consequence of heating concentrated at the footpoints of the loop, but constant in time.

 }
\maketitle

\section{Introduction}
Recent observations of the solar transition region and corona, especially with the Solar and Heliospheric Observatory (SOHO) and the Transition Region And Coronal Explorer (TRACE), have shown that magnetically closed structures in the upper solar atmosphere, commonly referred to as coronal loops, exhibit intrinsically dynamic behavior.
Even in quiescent, non-flaring conditions, loops show strong temporal variability of emission in UV spectral lines and substantial plasma flows.
An overview of observations of the temporal variability of active region loops with the Coronal Diagnostic Spectrometer (CDS) is given by \cite{Kjeldseth-Moe+Brekke1998SP}. They report significant changes of coronal loops over a period of one hour, in particular seen in emission lines in the temperature range between $T=1 - 5 \cdot 10^5$\,K. This variability is accompanied by large Doppler shifts, typically around $v = 50 - 100$\,km/s.
Recent observations with CDS and the Extreme ultraviolet Imaging Telescope (EIT) with high temporal cadence \citep[Fredvik 2002, private communication,][]{DeGroof+al2004AA} furthermore reveal spatially localized brightenings in coronal loops,
moving rapidly down towards the footpoints of the loops.
The fact that coronal loops can undergo rapid evacuation has been known for decades: \citet{Levine+Withbroe1977SP}, e.g., report \emph{Skylab} spectroscopic observations, compatible with ``dramatic evacuation'' of active region loops triggered by rapid, radiation dominated cooling.
A detailed study of ``catastrophic cooling'' and evacuation of quiescent coronal loops observed with the TRACE instrument is presented by \cite{Schrijver2001SP}. He analyzes image sequences taken in different spectral passbands and finds that loop evacuation occurs frequently after plasma in the upper parts of the loops has cooled to transition region or lower temperatures. The cooling process is often accompanied by emission in Ly$_\alpha$ and \ion{C}{4} (154.8\,nm), developing initially near the loop top. Thereafter, cool plasma is observed to slide down on both sides of the loop, forming clumps which move with velocities of up to 100\,km/s. The downward acceleration of these plasma clumps as inferred from these observations is significantly less than the gravitational acceleration on the solar surface.
According to the observations of \cite{Schrijver2001SP}, this process of dramatic cooling and evacuation is a rather common one.
Further observational evidence of ``blobs'' of plasma falling down towards the solar surface along magnetic field lines is presented by \cite{DeGroof+al2004AA}, based on high cadence time series of simultaneous EIT (30.4\,nm) and Big Bear $H_\alpha$ data.

In this paper, we present numerical models of coronal loops which exhibit a wide range of dynamics using a very simple heating function that is exponentially decreasing with height, but \emph{constant} in time. A key feature of these models is the recurrent formation of plasma condensations, followed by loop evacuation, as described in \citet[][referred to as Paper \rom{1} hereafter]{Mueller+al2003a}, which offers a unifying explanation for different aspects of recent observations.

\section{\label{sec_num}Numerical model}
In this work, we use the same numerical model as in Paper \rom{1}, and the reader is referred to \citet{Mueller+al2003a}  and \citet{Hansteen1993ApJ} for details. Our code solves the one-dimensional time-dependent hydrodynamic equations for mass, momentum and energy conservation, coupled with the ionization rate equations for several elements and self-consistent radiative losses. The plasma is assumed to be effectively thin, and the radiative losses are due to collisional excitation of the various ions comprising the plasma, in addition to thermal bremsstrahlung. Thermal conduction, radiative losses and a coronal heating term are included in the energy equation.
In the radiative losses the elements hydrogen, helium, carbon, oxygen, silicon, neon, and iron are included.
While some of the metals are treated by assuming ionization equilibrium and then deriving an \emph{a priori} radiative loss curve as a function of electron temperature, radiative losses from the ions specifically mentioned in this study, i.e. losses from hydrogen, helium, carbon and oxygen, are computed consistently with full time-dependent rate equations. 
We consider a loop of low-$\beta$ plasma and assume that the loop has a constant cross section.

\subsection{\label{sec:heat}Loop heating}
\begin{figure}
\resizebox{\hsize}{!}{
\includegraphics{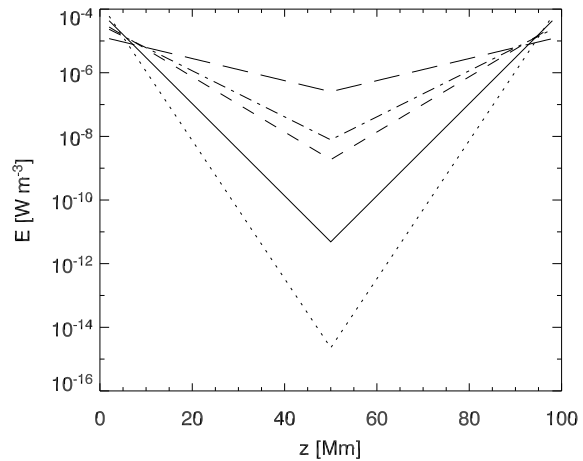}
}
\caption{\label{fig1}Prescribed heating rate for different values of the damping length: $H_m = 2$\,Mm (\emph{dotted}),  $H_m = 3$\,Mm (\emph{solid}),  $H_m = 5$\,Mm (\emph{dashed}), and  $H_m = 6$\,Mm (\emph{dash-dotted}), and $H_m = 12.5$\,Mm (\emph{long dashes}, heating function for static initial model). The total heat input into the loop is the same for all cases.}
\end{figure}

We parameterize the energy input into
the coronal loop by specifying the energy flux at the footpoints
of the loop, $F_{m0}$, and assuming a mechanical energy flux that is
constant up to a height $z_1$ and then decreases exponentially for $z \ge z_1$ as 
\be
\label{eq:heating}
F_m(z) = F_{m0}\exp[-(z-z_1)/H_m] 
\ee
with a damping length $H_m$.
In the models presented below, we will vary $H_m$ between 2 and 12.5\,Mm for a loop of 100\,Mm length.
For the mechanical energy flux we use $F_{m0} =
c \cdot 150$~W/m$^2$ with the normalization constant $c = 1/(1 - \exp[-(L/2-z_1)/H_m])$
and set $z_1 =1.75$\,Mm.
The normalization constant ensures that the total energy input into the loop is the same, irrespective of the damping length $H_m$.
The heating rate, i.e. the energy deposition per unit time and unit
volume, is given by the divergence of the energy flux, $Q_m = - \nabla F_m$. Fig.~\ref{fig1} displays graphs of the heating rate for different values of the damping length, $H_m$.
With the damping length $H_m$ we can control whether the heating is concentrated near the footpoints or more evenly distributed along the loop.

An exponentially decaying heating function was first suggested by \cite{Serio+al1981ApJ} and
seems to be supported by recent observations
\citep{Aschwanden+al2000ApJ541,Aschwanden+al2001ApJ550} as well as by numerical simulations of \citep{Gudiksen+Nordlund2002ApJL}.

\subsection{Initial state}
\begin{figure}
\resizebox{\hsize}{!}{
\includegraphics{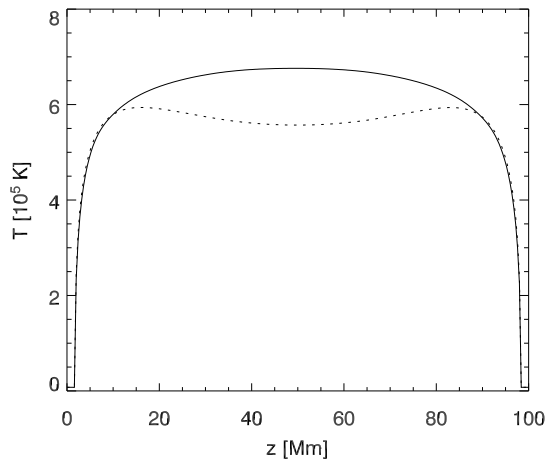}
}
\caption{\label{fig_init}Temperature along the coronal loop. Initial state (\emph{solid line}, $H_m = 12.5$\,Mm) and stable solution for $H_m = 6$\,Mm (\emph{dotted line}).}
\end{figure}

The coronal loop model studied here has a total length of 100\,Mm, composed of a semicircular arch of 98\,Mm length and a vertical stretch of 1\,Mm length at each end.
A static initial state is obtained by prescribing a large energy dissipation length of $H_m = 12.5$\,Mm, which results in a loop apex temperature of $T = 6.8 \cdot 10^5$\,K. The temperature along the loop of the initial state is plotted in Fig.~\ref{fig_init}.

\section{\label{sec_res}Plasma condensation due to thermal instability}
For very short damping lengths of the heating function solutions with a hot loop apex may no longer exist due to the insufficient energy supply to the center, i.e. the top of the loop, as shown by \cite{Antiochos+al1999ApJ}, \cite{Karpen+al2001ApJ} and \cite{Mueller+al2003a}.
In this case, a thermal instability occurs and leads to a runaway cooling process, also called \emph{catastrophic cooling}, accompanied by plasma condensation (cf.~Paper \rom{1} for details).
Unless this condensation region is gravitationally supported, for example by means of a dip in the magnetic field lines, to maintain a stable prominence-like state, such a configuration is unstable and the dense condensation region eventually moves down the loop legs and drains through the footpoints.
The depleted loop then reheats quickly as its heat capacity is very low (at this stage there is much less mass in the loop but the heating remains constant) and is filled again by chromospheric evaporation.
Exactly how this cycle of plasma condensation, draining, and chromospheric evaporation is realized depends strongly on the spatial dependence of the energy deposition.

In Paper \rom{1}, we studied the physical processes leading to this evaporation--condensation cycle  and its application to small ($L = 10$\,Mm) transition region loops, which can just barely be spatially resolved with the currently available instruments. This work focuses on plasma condensations in longer coronal loops ($L = 100$\,Mm), where the same process can induce significantly stronger flows and greater variations in the spectral signature, due to the longer acceleration phase along the loop. More generally, the aim of this paper is also to work out the different types of loop evolution that result from different damping lengths of the heating function.

It is interesting to note that a time-dependent evolution for
time-independent heating over short damping lengths has already been
described in a different context by \cite{Hearn+al1983AA} and \cite{Korevaar+Hearn1989AA}. However they applied their results not to solar coronal loops, but to open coronal regions surrounding hot stars.

Cyclic evolution of coronal loops was studied for the first time by
\cite{Kuin+Martens1982AA}. In their semi-analytical model, they treated
the coronal loop as an integrated system, coupled to the underlying
chromosphere. A comparison of their work with our hydrodynamical simulations is given in Paper \rom{1}.

\section{Results}
\subsection{\label{sec:types}Different types of loop evolution}
For large damping lengths ($H_m \geq 6$\,Mm) and a prescribed energy flux as described in Sect.~\ref{sec:heat}, a stable, static loop solution is attained.
Fig.~\ref{fig_init} shows the temperature along the loop for $H_m = 6$\,Mm, which has a mean temperature of $\langle T \rangle = 5.3 \cdot 10^5$\,K.\footnote{Throughout this paper, the mean values are defined as the
average quantities over the region of the loop which lies above the
transition region, bounded by the points where the temperature crosses
$T= 10^5$\,K in both loop legs (the exact choice of this cut-off value
does not significantly influence the results and could be set to any
temperature $T \gtrsim 2\cdot 10^4$\,K).}
For shorter damping lengths, when the heating is more concentrated at the footpoints, the loop loses its thermal equilibrium and exhibits a dynamic evolution. 
\begin{figure}
\resizebox{\hsize}{!}{
\includegraphics{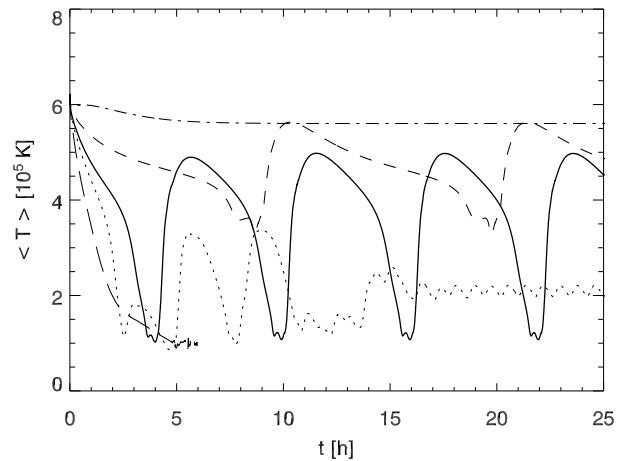}
}
\caption{\label{fig2}Evolution of mean temperature, $\langle T \rangle$, as a function of time, for four different damping lengths of the heating function: $H_m = 2$\,Mm (\emph{dotted}), $H_m = 3$\,Mm (\emph{solid}), $H_m = 5$\,Mm (\emph{dashed}), $H_m = 6$\,Mm (\emph{dash-dotted}). Cf.\ Sect.\ \ref{sec:types}. For comparison, the \emph{long-dashed} line shows $\langle T \rangle (t)$ for a loop model where the heating is switched off at $t=0$.}
\end{figure}
Fig.~\ref{fig2} shows the mean temperature, $\langle T \rangle$, of a $L=100$\,Mm loop as a function of time for damping lengths of $H_m = 2, 3, 5,$ and 6\,Mm.

We find that for 2.5\,Mm $< H_m <$ 6\,Mm, the loop shows a periodic variation of $\langle T \rangle$ due to the evaporation--condensation cycle as described in Paper \rom{1}. For even shorter damping lengths ($H_m \leq 2.5$\,Mm), the evolution of  $\langle T \rangle$ is irregular and shows intermittency of hot phases and strongly fluctuating cool phases. This type of intermittent behavior is well-known from chaotic non-linear systems.

For comparison, Fig.~\ref{fig2} displays also $\langle T \rangle (t)$ for a loop model where the heating is switched off at $t=0$ (\emph{long-dashed line}). In this case, the loop plasma simply drains on both sides of the loop with flow speeds of $v < 15$\,km/s, and the loop cools down to chromospheric temperatures without any plasma condensation forming.

Let us examine the different types of dynamic solutions in more detail to see which phenomena accompany the condensation process. In Fig.~\ref{fig3}, we plot space-time diagrams of the loop temperature, $T (z,t)$, for $H_m = 2, 3, 5$\,Mm. The \emph{left} and the \emph{center} plot show two different kinds of recurrent formation of plasma condensations:

\begin{figure*}
\begin{center}
\includegraphics[height=24cm]{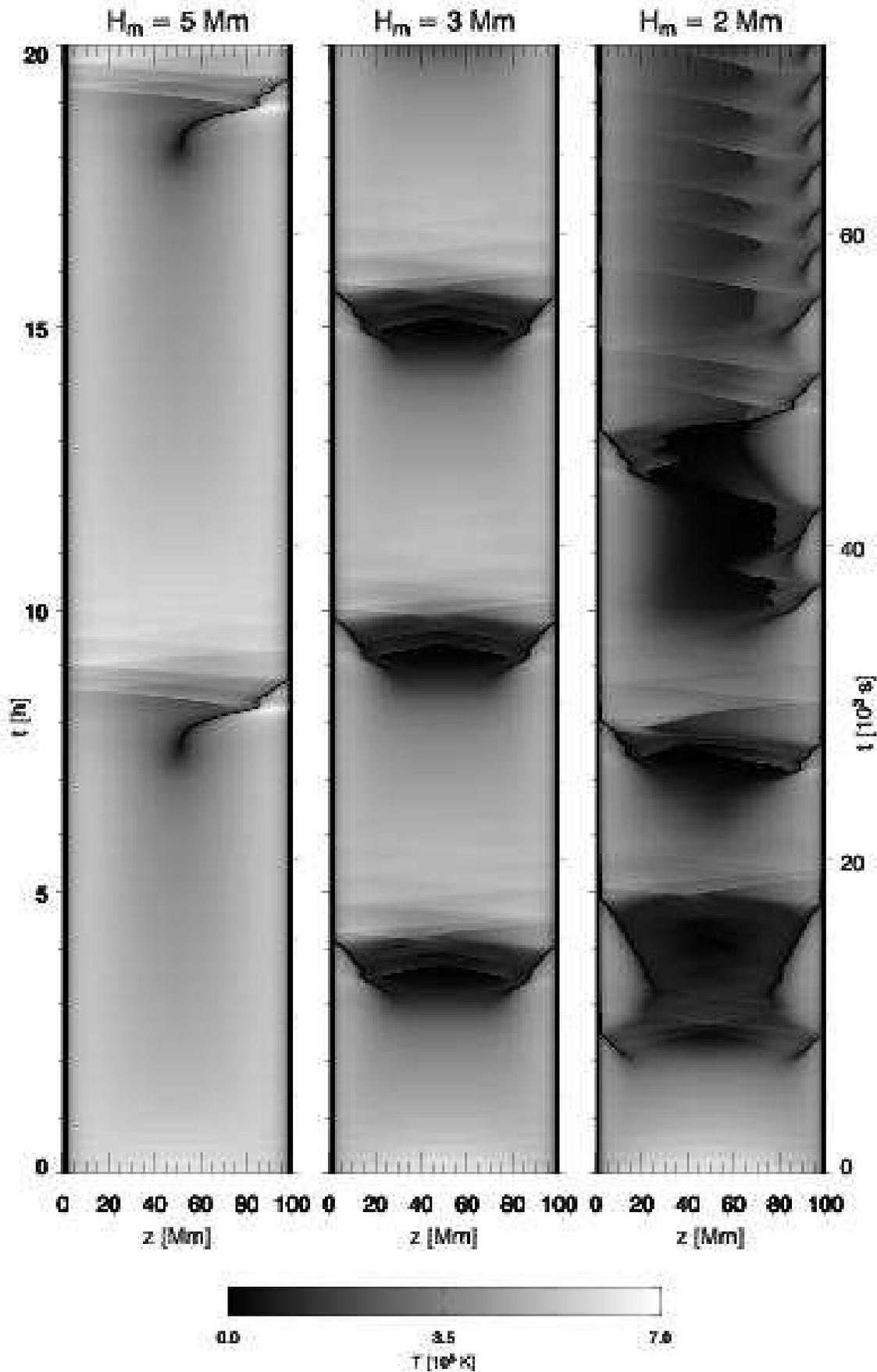}
\end{center}
\caption{\label{fig3}Evolution of the temperature along the loop, $T (z,t)$, for three different damping lengths of the heating function: $H_m = 5$\,Mm (\emph{left}), $H_m = 3$\,Mm (\emph{center}), $H_m = 2$\,Mm (\emph{right}). The loop footpoints are at $z=$0 and 100\,Mm, the apex is at $z=$50\,Mm.}
\end{figure*}

In the first case ($H_m = 5$\,Mm), one condensation region forms at the loop apex and is then accelerated on its way down, resulting in flow velocities in the wake of the falling plasma blob of up to $v \approx 100$\,km/s. When this condensation region encounters the transition region near the loop footpoint, it is strongly decelerated by the pressure gradient of the underlying plasma and the velocity profile forms a shock front. As the compression is approximately adiabatic, this leads to a transient heating of the transition region plasma.

The direction in which the blob starts to move is decided by small asymmetries of the pressure around the loop apex. A small increase of the deposited energy in one loop leg (e.g.~$1\%$) is sufficient to trigger a motion of the blob in the opposite direction.
In Sect.~\ref{sec:shock}, we will discuss the velocity profiles of the flow in more detail and point out a possible connection to the recent observations of falling plasma blobs by \cite{DeGroof+al2004AA}.

The \emph{center} plot of Fig.~\ref{fig3} shows the second type of recurrent condensations which occurs if the damping length is slightly reduced with respect to the first case ($H_m = 3$\,Mm). Here, two condensation regions form simultaneously and then drain down both loop legs.
We note that, in agreement with the results obtained in Paper \rom{1} for short loops, the period of the condensation cycle decreases with decreasing damping length as the loss of equilibrium due to insufficient heating of the upper part of the loop occurs sooner.

The \emph{right} panel of Fig.~\ref{fig3} shows the most complex evolution of this set of numerical experiments: As the heating is even more concentrated towards the footpoints, the evolution of temperature along the loop with time reflects the persistent battle between loop heating and radiative cooling:
 The loop first cools down from its initial state to $T \approx 1.1\cdot 10^5$\,K and forms two condensation regions at $t = 8\,000$\,s ($\approx 2.2$\,h). After these have drained, the loop starts reheating. Due to the concentration of the heating to low heights, however, not enough energy is deposited in the upper part of the loop to prevent it from repeated radiative cooling and condensation at  $t = 12\,000$\,s. At  $t = 19\,000$\,s, the loop recovers from its catastrophic cooling and enters a quiet, warm phase, during which the flow speed does not exceed $v = 16$\,km/s. At  $t = 25\,000$\,s, a new instability sets in and leads to the formation of two new condensation regions whereupon flow speeds of up to $v = 95$\,km/s are reached. The reflections of the shock fronts meet near the loop apex and yield to a transient temperature increase there. At  $t = 30\,000$\,s, a new phase of evolution starts: small condensation regions are recurrently formed in one leg of the loop, but due to the footpoint-centered heating function, the loop does not reach temperatures of more than $T = 3.9 \cdot 10^5$\,K before collapsing again.
At $t = 54\,000$\,s, the loop enters a periodic phase where condensation regions are recurrently formed in the right loop leg. 

How long the different phases of loop evolution last is dependent on small variations in the radiative loss rate. We observed ``chaotic'' evolution of the loop for the entire duration of the longest simulation run ($2\cdot10^5$\,s) when including the non-equilibrium ionization of only hydrogen and helium, while the loop reached a periodic solution at $t = 54\,000$\,s (as described above) when also accounting for the non-equilibrium ionization of carbon and oxygen.

\subsection{\label{sec:phase}Classification of loop evolution}

\begin{figure}
\begin{center}
\includegraphics[width=7cm]{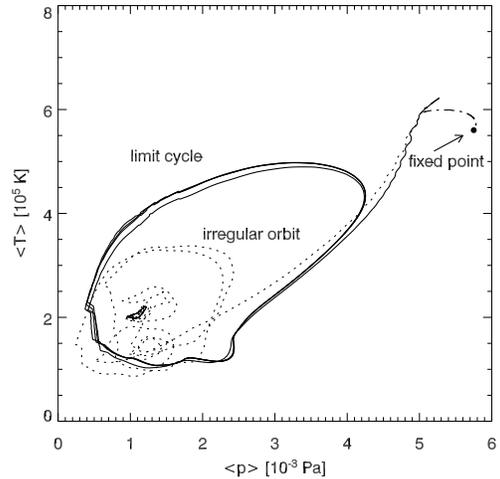}
\end{center}
\caption{\label{fig4} Mean temperature, $\langle T \rangle$, of the loop, as a function of mean pressure, $\langle p \rangle$, for a loop of total length $L = 100$\,Mm. \emph{Dotted}: $H_m$ = 2\,Mm, \emph{solid}: $H_m$ = 3\,Mm, \emph{dash-dotted}: $H_m$ = 6\,Mm.}
\end{figure}

One way of representing the temporal evolution of coronal loops is in terms of phase diagrams in $\langle p \rangle - \langle T \rangle$ space. Fig.~\ref{fig4} shows such a phase diagram for a static loop ($H_m = 6$\,Mm, \emph{dash-dotted}), a periodically condensing loop ($H_m = 3$\,Mm, \emph{solid}), and an irregular loop ($H_m = 2$\,Mm, \emph{dotted}).
We see that the stable loop approaches a fixed point in $\langle p \rangle - \langle T \rangle$ space, while the periodically condensing loop traces out limit cycles. On the other hand, the irregular loop exhibits a pattern that is composed of a multitude of small intersecting paths, occasionally interrupted by larger cycles corresponding to the temporarily stable phases of the loop.

\subsection{\label{sec:wing}Where in a coronal loop do condensation regions form?}

In the beginning of the cooling process, the evolution of the temperature as a function of loop length is very similar for the two cases of $H_m = 3$\,Mm and $H_m = 5$\,Mm. This raises the question why two condensation regions form in the wing of the loop in one case and only one central condensation region in the other case, where the heating is less concentrated towards the footpoints.

\subsubsection{\label{energy}Some consideration on the energetics}
In order to better understand the formation of the condensation region it
is also helpful to study the deposition and transport of the energy.
If the ratio of the damping length to the loop length is large, stable coronal loops reach the maximum temperature near the loop top. In contrast, stable loops with a smaller ratio of damping length to loop length reach the peak temperature well below the apex and have a rather flat temperature profile in the central part of the loop.
For instance, a stable 100 Mm long loop with $H_m = 6$\,Mm reaches a maximum temperature of
$5.9\cdot 10^5$\,K some 16\,Mm above its footpoints, while the central 80\,Mm,
i.e.\ most of the loop, show only a 6\% change in temperature, with a
local temperature minimum of $5.6\cdot 10^5$\,K at the apex. 
In this model, the loop top is not predominately heated by the mechanical
heating as defined in Eq.\ (\ref{eq:heating}).
The rapid exponential decrease of the heating can sustain high coronal
temperatures only up to some 10\,Mm height above each footpoint.
Above that height the plasma is mainly heated by heat conduction.
In an equilibrium situation this leads to a temperature dip at the loop
apex.
As the heat conduction is efficient at high temperatures the resulting
temperature profile is rather flat in the upper part of the loop.

If the heating is more and more concentrated to the footpoints (by
decreasing $H_m$), the peak temperature becomes smaller and occurs at lower
heights.
This reduces the heat input through heat conduction into the upper part of
the loop, and finally the heat conduction can no longer balance the
radiative losses and catastrophic cooling sets in.
This clarifies why the catastrophic drop in
temperature can set in over a very wide range of the loop, basically in the
whole region between the temperature maxima (cf.\ \emph{middle} and \emph{right} panel of Fig.~\ref{fig3} and \emph{upper left} panel of Fig.~\ref{h3ana}).

\subsubsection{Off-center formation of condensation regions}

\begin{figure*}
\begin{center}
\resizebox{0.7\hsize}{!}{
\includegraphics{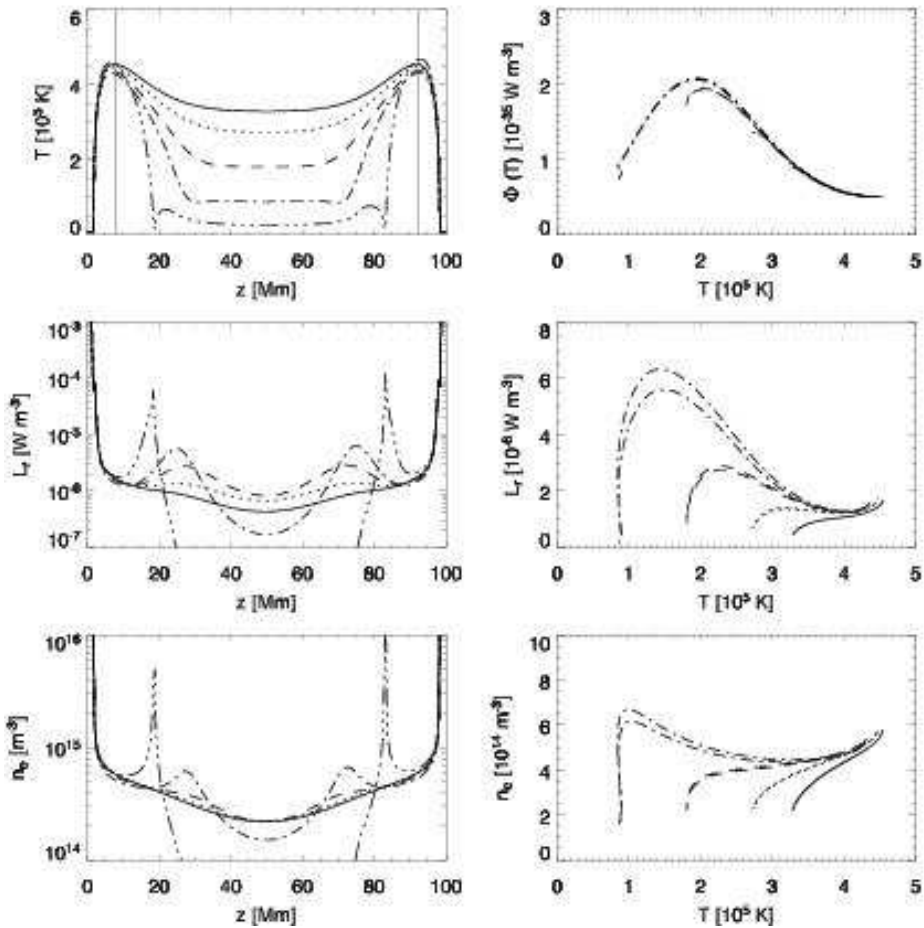}
}
\caption{\label{h3ana}Formation of two simultaneous, lateral condensation regions for $H_m = 3$\,Mm. \emph{Left} panels, as functions of loop length: loop temperature (\emph{top}), total radiative losses (\emph{middle}), electron density (\emph{bottom}). \emph{Right} panels, as functions of loop temperature: radiative loss function (\emph{top}), total radiative losses (\emph{middle}), electron density (\emph{bottom}). The following timesteps are plotted: $t=30\,000\,s$ (\emph{solid}), $t=31\,000\,s$ (\emph{dotted}), $t=32\,000\,s$ (\emph{dashed}), $t=33\,000\,s$ (\emph{dash-dotted}), $t=34\,000\,s$ (\emph{dash-dot-dotted}). The \emph{right} panels display only data of the central part of the loop, between the two vertical lines shown in the \emph{upper left} panel, for the first 4 timesteps. In some cases (\emph{dashed} and \emph{dash-dotted} curves in the \emph{middle} and \emph{lower right} panels) two branches are seen because of a slightly different evolution of the two loop legs.}
\end{center}
\end{figure*}

The \emph{upper left} panel of Fig.~\ref{h3ana} shows the temperature profiles of the $H_m = 3$\,Mm loop for five different timesteps. In the first timestep, $t = 30\,000$\,s, the loop is already in the cooling phase, and the temperature decreases with time throughout the central part of the loop. The cooling is dominated by radiation, with total radiative losses of $L_r = \Phi(T)\cdot n_e \cdot n_H \approx \Phi(T)\cdot n_e^2$.
The radiative loss function, $\Phi(T)$, is determined by the ionization (non)equilibrium of the model atoms included in the calculation and is therefore time-dependent. However, if the ionization of the loop plasma does not depart too strongly from equilibrium, the radiative losses peak around $T\approx 2 \cdot 10^5$\,K, as shown in the \emph{upper right} panel of Fig.~\ref{h3ana}. This means that plasma of a given electron density cools more efficiently at, e.g., $T = 2 \cdot 10^5$\,K than at $T = 4 \cdot 10^5$\,K. On the other hand, the density enters quadratically into in the radiative loss function, so that a local density enhancement anywhere leads to a strongly increased cooling.
If we now concentrate on the timestep $t = 31\,000$\,s (\emph{dotted line}) and compare the different panels on the \emph{left} side of Fig.~\ref{h3ana}, we see that the total radiative losses have developed two local maxima in the wings of the loop, which subsequently lead to local density maxima at  $t = 33\,000$\,s (\emph{dash-dotted line}). This initiates the formation of two condensation regions, as seen by the drastic density increase at later timesteps. At the earlier time $t = 31\,000$\,s, however, these local maxima are not yet accompanied by local density maxima and arise at a temperature of $T \approx 3 \cdot 10^5$\,K (\emph{right center} panel of Fig.~\ref{h3ana}), which is not the location of the maximum of the radiative loss function, $\Phi(T)$. The total radiative losses, $L_r \approx \Phi(T)\cdot n_e^2$, however, peak here and hence lead to the formation of lateral condensation regions. The \emph{lower right} panel of Fig.~\ref{h3ana} shows the strong density decrease towards the cooler center of the loop which prevents the formation of a central condensation region.

\subsubsection{Formation of central condensation regions}

\begin{figure*}
\begin{center}
\resizebox{0.7\hsize}{!}{
\includegraphics{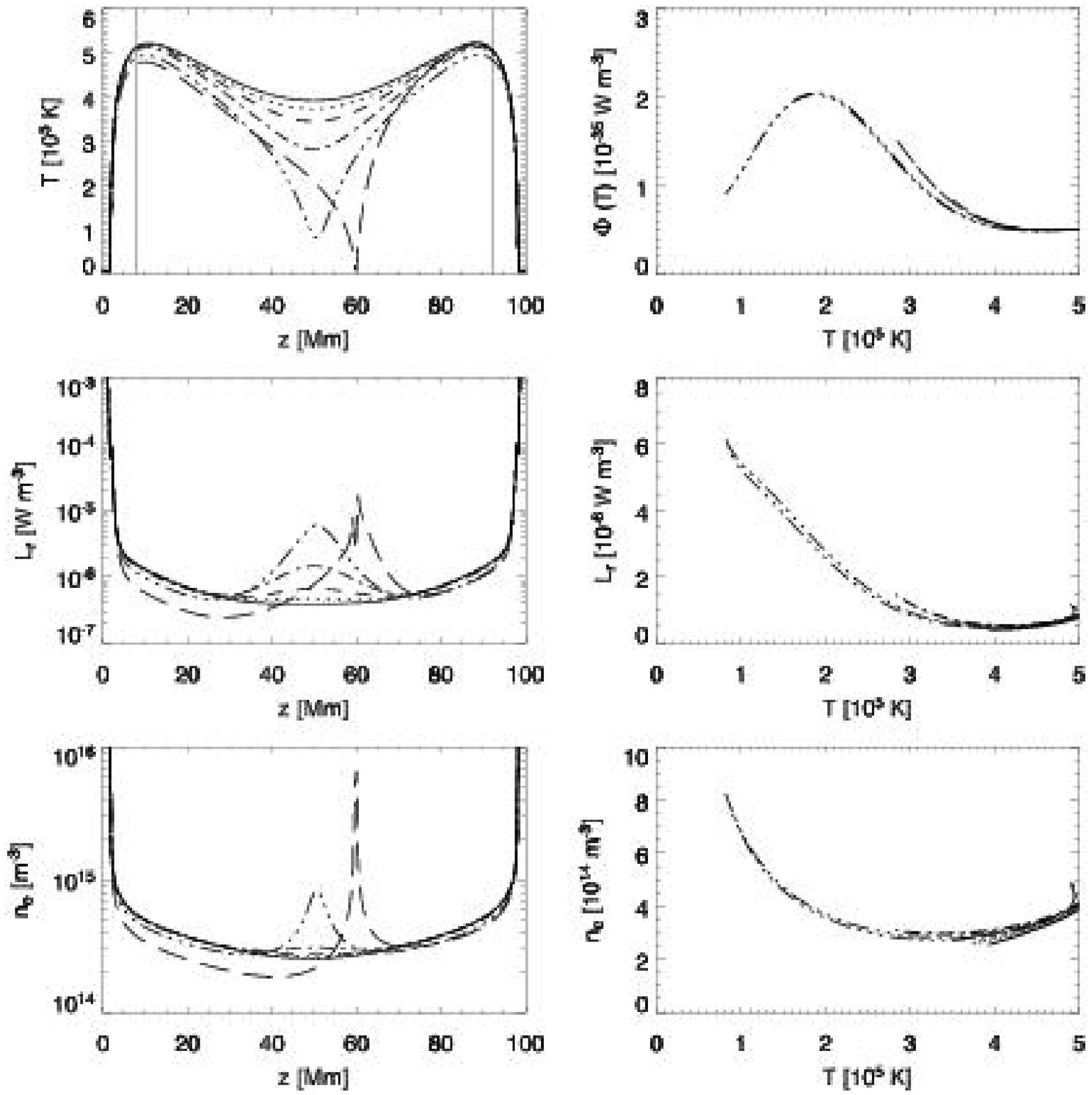}
}
\caption{\label{h5ana}Formation of a central condensation region for $H_m = 5$\,Mm. \emph{Left} panels, as functions of loop length: loop temperature (\emph{top}), total radiative losses (\emph{middle}), electron density (\emph{bottom}). \emph{Right} panels, as functions of loop temperature: radiative loss function (\emph{top}), total radiative losses (\emph{middle}), electron density (\emph{bottom}). The following timesteps are plotted: $t=20\,000\,s$ (\emph{solid}), $t=22\,000\,s$ (\emph{dotted}), $t=24\,000\,s$ (\emph{dashed}), $t=26\,000\,s$ (\emph{dash-dotted}), $t=28\,000\,s$ (\emph{dash-dot-dotted}), $t=30\,000\,s$ (\emph{long dashes}). The \emph{right} panels display only data of the central part of the loop, between the two vertical lines shown in the \emph{upper left} panel, for the first 5 timesteps.}
\end{center}
\end{figure*}

We now compare the results obtained for a damping length of $H_m = 3$\,Mm with those for  $H_m = 5$\,Mm.
In the \emph{upper left} panel of Fig.~\ref{h5ana}, the temperature profiles of the $H_m = 5$\,Mm loop for six different timesteps are plotted. As in the previous case, the loop is in its cooling phase, with a local temperature minimum at the loop apex, and cools fastest around the apex. In contrast to the $H_m = 3$\,Mm case, however, the density gradient, $\partial n_e / \partial z$, near the apex is significantly shallower than in the previous case. This is due to the larger damping length which means that a larger fraction of the energy is dissipated higher up in the loop. \footnote{The larger damping length also results in a higher temperature in general. When the cooling phase sets in, the $H_m = 5$\,Mm loop has a maximum temperature of $T=5.8\cdot 10^5$\,K compared to $T=5.2\cdot 10^5$\,K  for the  $H_m = 3$\,Mm loop (cf.\ also the mean temperatures plotted in Fig.~\ref{fig2}).} Therefore, a local maximum of the total radiative losses forms at the loop apex at $t = 22\,000$\,s (\emph{middle left} panel, \emph{dotted line}). The \emph{middle right} panel displays the total radiative losses, $L_r$, as a function of temperature, and it is seen that for $T < 4\cdot 10^5$\,K, $L_r(T)$ increases monotonically with decreasing temperature. Consequently, a local density maximum forms at the loop apex at $t = 26\,000$\,s (\emph{lower left} panel, \emph{dash-dotted line}) and evolves by catastrophic cooling into a condensation region.

\subsection{\label{sec:shock}Formation of a shock front}

\begin{figure}
\resizebox{\hsize}{!}{
\includegraphics[width=13cm]{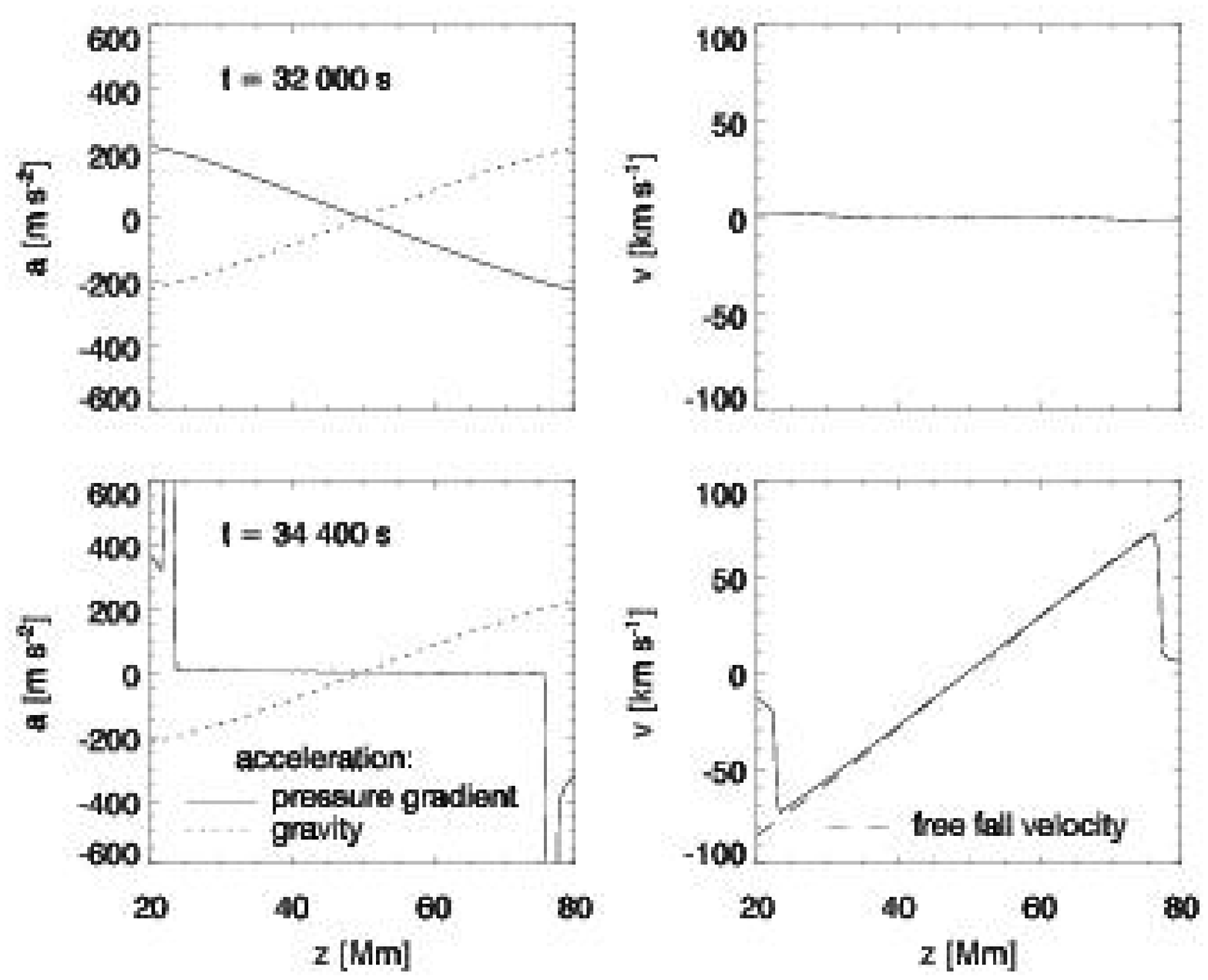}
}
\caption{\label{fig7}Formation of a shock front for $H_m = 3$\,Mm. Gravitational acceleration, $g_\| (z)$ (\emph{dotted}), and acceleration due to the pressure gradient, $\nabla p (z)/\rho(z)$ (\emph{solid}) for $t = 32\,000$\,s (\emph{top} row) and $t = 34\,400$\,s (\emph{bottom} row). In the \emph{lower right} plot, the free-fall velocity profile is indicated by a \emph{dashed line}.}
\end{figure}

In a hydrostatic configuration, the gravitational force acting on the plasma is balanced by the pressure gradient. As an illustration of a temporarily static phase of the coronal loop with $H_m = 3$\,Mm, we plot in Fig.~\ref{fig7}  (\emph{upper left} panel) the component of the gravitational acceleration parallel to the loop, $g_\| (z)$,  and the acceleration due to the pressure gradient, $\nabla p (z)/\rho(z)$, along the coronal part of the loop at $t = 32\,000$\,s. It is seen that these two quantities compensate each other, and due to this equilibrium, the plasma in the loop is nearly static (the velocity is displayed in the \emph{upper right} panel).

At  $t = 34\,400$\,s, however, a loss of equilibrium has occurred and the gravitational force is no longer balanced by the pressure gradient (\emph{lower left} panel): In the central part of the loop, the pressure gradient has dropped to very small values, while close to the footpoints, it is more than an order of magnitude higher than the gravitational acceleration.
The reason why $\nabla p$ becomes much smaller in the central part of the loop than in hydrostatic equilibrium is the drastic decrease of the temperature  (cf.~Fig.~\ref{h3ana}, \emph{upper left} panel) which causes a strong decrease in the pressure.
This explains the velocity profile seen in the \emph{lower right} panel: In the central part of the loop, the plasma is accelerated to velocities very close to the free-fall speed, indicated by the \emph{dashed line}, and then strongly decelerated in the lower parts of the loop, resulting in a characteristic shock profile (Positive values of $v$ denote a flow in the positive $z$-direction. A downflow in the left loop leg ($z < 50$\,Mm) is thus characterized by velocities $v < 0$, while a downflow in the right loop leg ($z > 50$\,Mm) has velocities $v > 0$.).

\subsection{\label{sec:vel_accel}Velocity profiles and acceleration of the condensation region}

In order to compare our results with the velocities and accelerations deduced from observations of ``moving blobs'' in coronal loops, we concentrate in this section on the falling condensation region around $t = 30\,000$\,s of the simulation run with $H_m = 5$\,Mm.

\begin{figure}
\resizebox{\hsize}{!}{
\includegraphics[width=7cm]{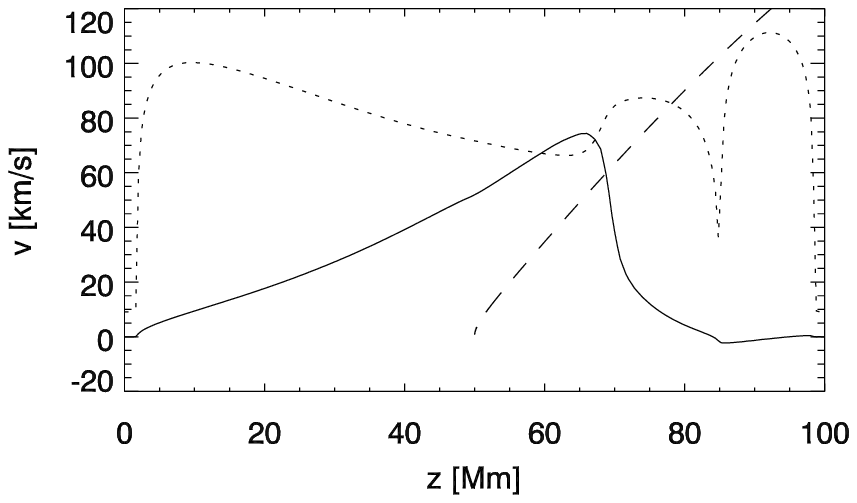}
}
\caption{\label{fig7a}Comparison between the velocity profile for $H_m = 5$\,Mm at $t = 31\,200$\,s (\emph{solid line}) with the free-fall velocity from the loop apex at 50\,Mm to the right footpoint at 100\,Mm (\emph{dashed}) and the local sound speed at $t = 31\,200$\,s (\emph{dotted line}).}
\end{figure}

Fig.~\ref{fig7a} shows the velocity profile for $t = 31\,200$\,s (\emph{solid line}) and a velocity profile corresponding to a free-fall of a test particle along the loop, which has a height of $h = 33.2$\,Mm (\emph{dashed line}). The maximum of the free-fall velocity is $v_{\rm max} = \sqrt{2 g h} = 135$\,km/s. Comparing the free-fall velocity profile with the velocity in the right half of the loop, it is seen that from $z = 50 - 70$\,Mm the flow is faster than free fall, while it is slower for $z = 70 - 100$\,Mm. The \emph{dotted line} displays the local sound speed; from $z = 60 - 65$\,Mm, the flow is supersonic. The fact that $v_{\rm apex} = 50$\,km/s immediately shows that there is a force acting on the loop plasma, which turns out to be the pressure force originating from the pressure difference between the wake of the moving condensation region and the rest of the loop behind the condensation region, which is located at $z = 85$\,Mm in this plot.

\begin{figure}
\begin{center}
\includegraphics[width=8cm]{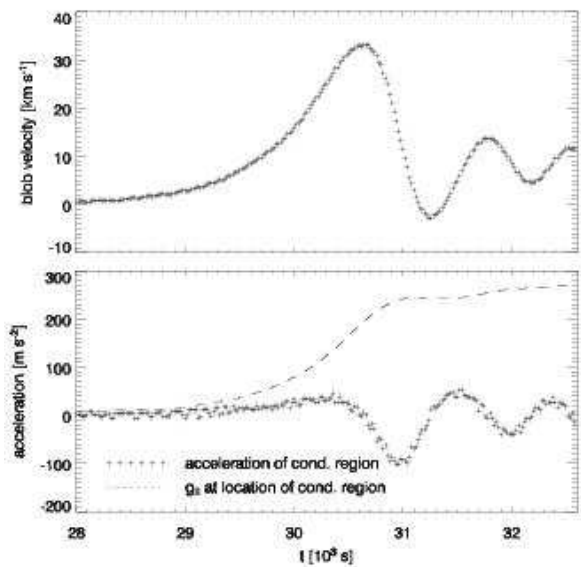}
\end{center}
\caption{\label{fig6}Velocity (\emph{top}) and acceleration (\emph{bottom}) of the condensation region for $H_m = 5$\,Mm. The blob is accelerated by gravity and then slowed down by the pressure of the compressed transition region plasma underneath. The \emph{dashed line} in the lower panel shows the effective gravitational acceleration, $g_\|$, at the respective position of the blob.}
\end{figure}

In Fig.~\ref{fig6} we plot the velocity and acceleration of the center of the condensation region as a function of time. For this purpose, the condensation region is defined as the interval in which the temperature drops below $T = 10^5$\,K (alternatively, a threshold for the density or the radiative losses could be used).
The increasing velocity in the left half of the \emph{upper} panel shows how the blob is being accelerated up to $v = 33$\,km/s. 
It can be seen in the \emph{lower} panel that for $t < 29\,500$\,s, the acceleration is only a little smaller than $g_\|$, i.e. the free-fall case. After $t = 30\,600$\,s, however, the pressure of the compressed plasma underneath has become so large (cf.\ Sect.~\ref{sec:types}) that the blob is now effectively \emph{decelerated}.
At $t = 31\,140$\,s, the blob stops and even bounces 1\,Mm upwards before falling again. After a second deceleration phase the blob drains through the loop's footpoint at $t = 32\,840$\,s.
The maximal acceleration of the blob during its fall is $a= 54 $\,m/s$^2$.

It has to be stressed that our model is one-dimensional, so that in reality the deceleration process may not be as vigorous as in the simulation described here.
If the magnetic field is weak, the enhanced pressure of a region of dense plasma will distort the magnetic field which can lead to a lateral expansion of the dense plasma and a storage of energy in the surrounding plasma and magnetic field \citep{Athay+Holzer1982ApJ}. \cite{Mackay+Galsgaard2001SP}, on the other hand, carried out two-dimensional simulations of the evolution of a density enhancement in a stratified atmosphere and find that a sufficiently strong magnetic field enables the density enhancement to maintain its shape as it falls, and indeed results in the dense blob rebounding several times.

The deceleration of the plasma blob in our model is caused by the same mechanism as proposed by \cite{Schrijver2001SP} and yields a blob acceleration  which is significantly lower than solar gravity and is consistent with the values of $a = 80 \pm 30$\,m/s$^2$ reported by \cite{Schrijver2001SP}.
The maximal \emph{blob} velocities obtained from our simulations are smaller than the maximal velocities of up to $100$\,km/s reported by \cite{Schrijver2001SP} and $60 - 110 $\,km/s \citep{DeGroof+al2004AA},
 while the maximal \emph{flow} velocities we obtain ($v_{\rm max} = 75$\,km/s for $H_m = 5$\,Mm, $v_{\rm max} = 74$\,km/s for $H_m = 3$\,Mm,  and $v_{\rm max} = 128$\,km/s for $H_m = 2$\,Mm) are of the same order. In our simulations, the highest flow speeds are reached in the wakes of the falling plasma blobs.
However, increasing the loop length results in a longer acceleration path, so that higher \emph{blob} velocities are obtained for loops longer than 100\,Mm. Further work on a comparison with observational data is in progress.

\subsection{\label{sec:spec_sig}Spectral signature of catastrophic cooling and downflows}

As our numerical code consistently solves the atomic rate equations for different atomic species, we can calculate the emission of a large number of coronal and transition region spectral lines including the effects of non-equilibrium ionisation. In this context, the emission in the lines of \ion{C}{4} (154.8\,nm) (formation temperature $T_f \approx 10^5$\,K) and \ion{O}{5} (63.0\,nm) ($T_f \approx 2.2 \cdot 10^5$\,K) is of particular interest since the 160\,nm passband filter of TRACE is dominated by \ion{C}{4} emission, and the \ion{O}{5} line is frequently observed with SOHO/CDS.
We analyze the simulation run with $H_m = 5$\,Mm and focus on the same period that was discussed in the previous section.

\begin{figure}
\begin{center}
\includegraphics[width=8cm]{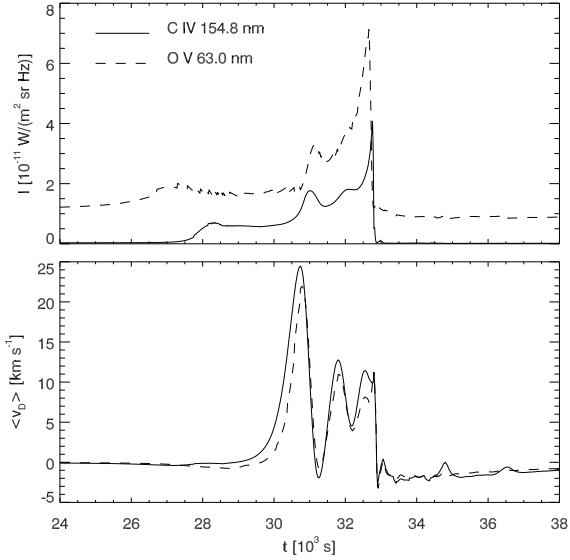}
\end{center}
\caption{\label{u_itot}Variation of total intensity (\emph{top}) and mean Doppler shift (\emph{bottom}) due to the falling condensation region, integrated over the right half of $H_m = 5$\,Mm loop. The \emph{solid line} displays \ion{C}{4} (154.8\,nm), the \emph{dashed line} \ion{O}{5} (63.0\,nm).
}
\end{figure}

In Fig.~\ref{u_itot} we plot the intensities and mean Doppler shifts, $\langle v_D \rangle$, for \ion{C}{4} (154.8\,nm) and \ion{O}{5} (63.0\,nm). Both quantities are integrated over the right half of the loop, excluding the footpoints, and the Doppler shifts are calculated as seen from above and converted to velocity units.

It is seen in the \emph{upper} panel that the blob brightens strongly in both spectral lines while falling, and reaches its maximal intensity shortly before draining through the footpoint.
The maximal Doppler shifts occur around $t = 30\,500$\,s, when the blob reaches its maximal velocity.
The maximal Doppler shifts are $\langle v_D \rangle = 25$\,km/s for \ion{C}{4} (154.8\,nm) and $\langle v_D \rangle = 22$\,km/s for
\ion{O}{5} (63.0\,nm).
Both lines are redshifted due to the blob's motion towards the solar surface.
Larger maximal Doppler shifts would result if no averaging over the entire right half of the loop was performed.

\begin{figure}
\resizebox{\hsize}{!}{
\includegraphics{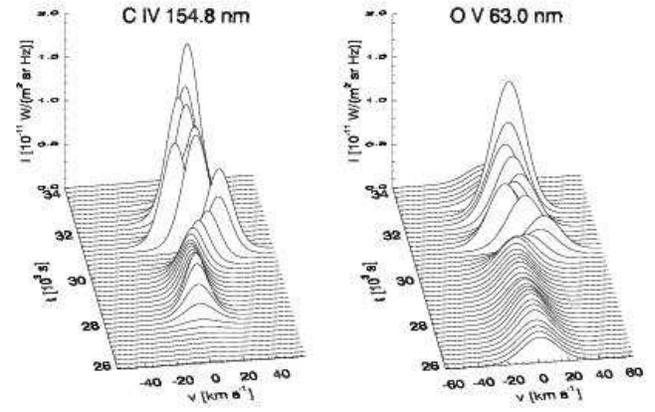}
}
\caption{\label{c+o_surf}Line profiles of \ion{C}{4} (154.8\,nm) and \ion{O}{5} (63.0\,nm) during the fall of the condensation region ($H_m = 5$\,Mm) as seen from above. The flow towards the solar surface results in a redshift of around 10\,km/s and the emission stops abruptly when the condensation region drains through the footpoint.}
\end{figure}

To visualize the variation of line shifts and intensity with time, we plot in Fig.~\ref{c+o_surf} line profiles of \ion{C}{4} (154.8\,nm) and \ion{O}{5} (63.0\,nm) for different points of time during the fall of the condensation region.
It is seen that the line profiles are redshifted as the blob falls while the line intensity increases (cf.~Fig.~\ref{u_itot}).

The Doppler shifts that would be measured in these lines will of course depend on the aspect angle that loop is viewed at. In order to calculate Doppler shifts which can directly be compared with, e.g., measurements with the CDS instrument on SOHO, one needs to consider not only the spatial resolution of the instrument, but also the finite temporal resolution due to the raster scan process. Further work is being carried out which will address these issues and present simulated raster scans of dynamic loops in different spectral lines. 

\begin{figure*}
\begin{center}
\includegraphics[width=13cm]{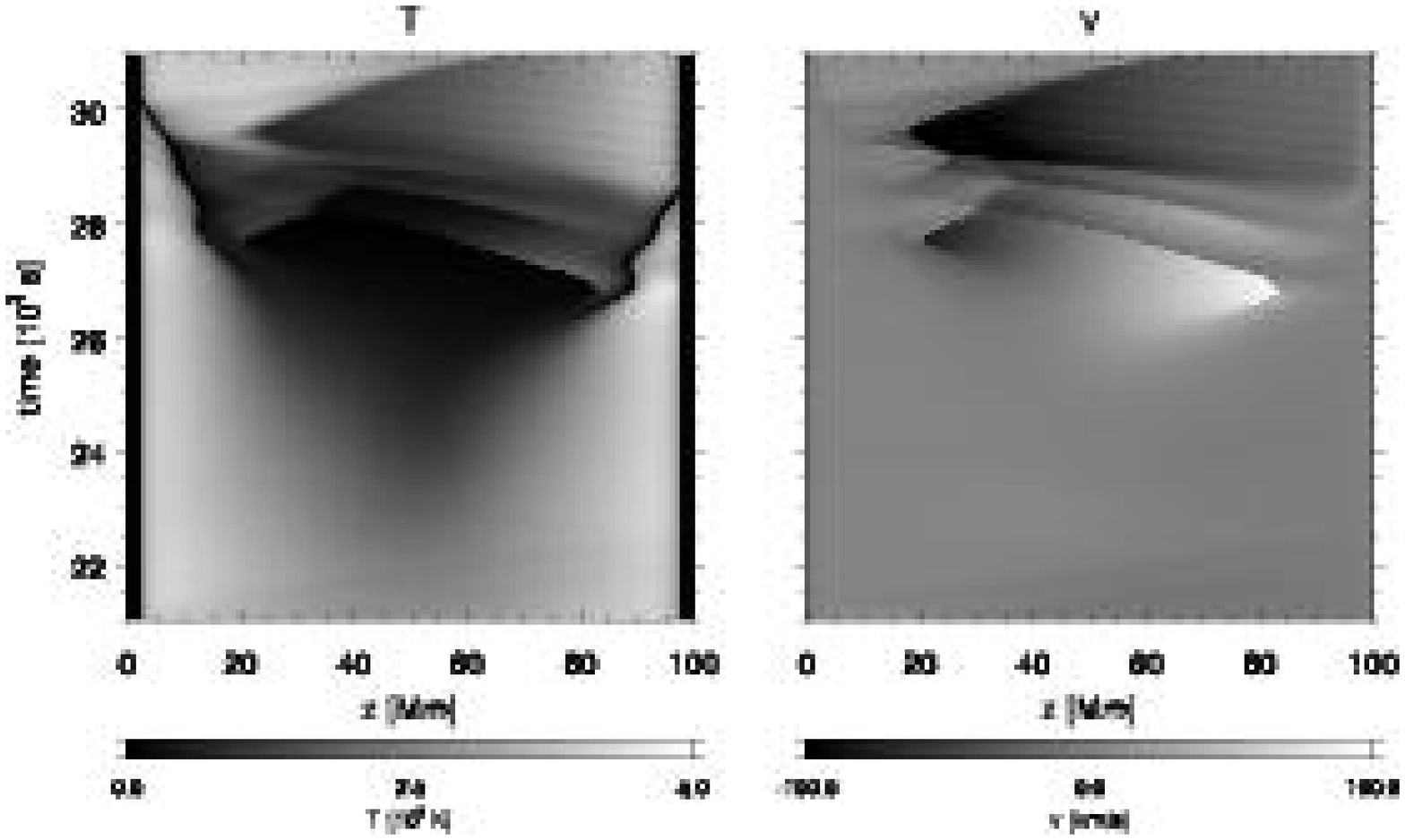}

\vspace{3mm}

\includegraphics[width=13cm]{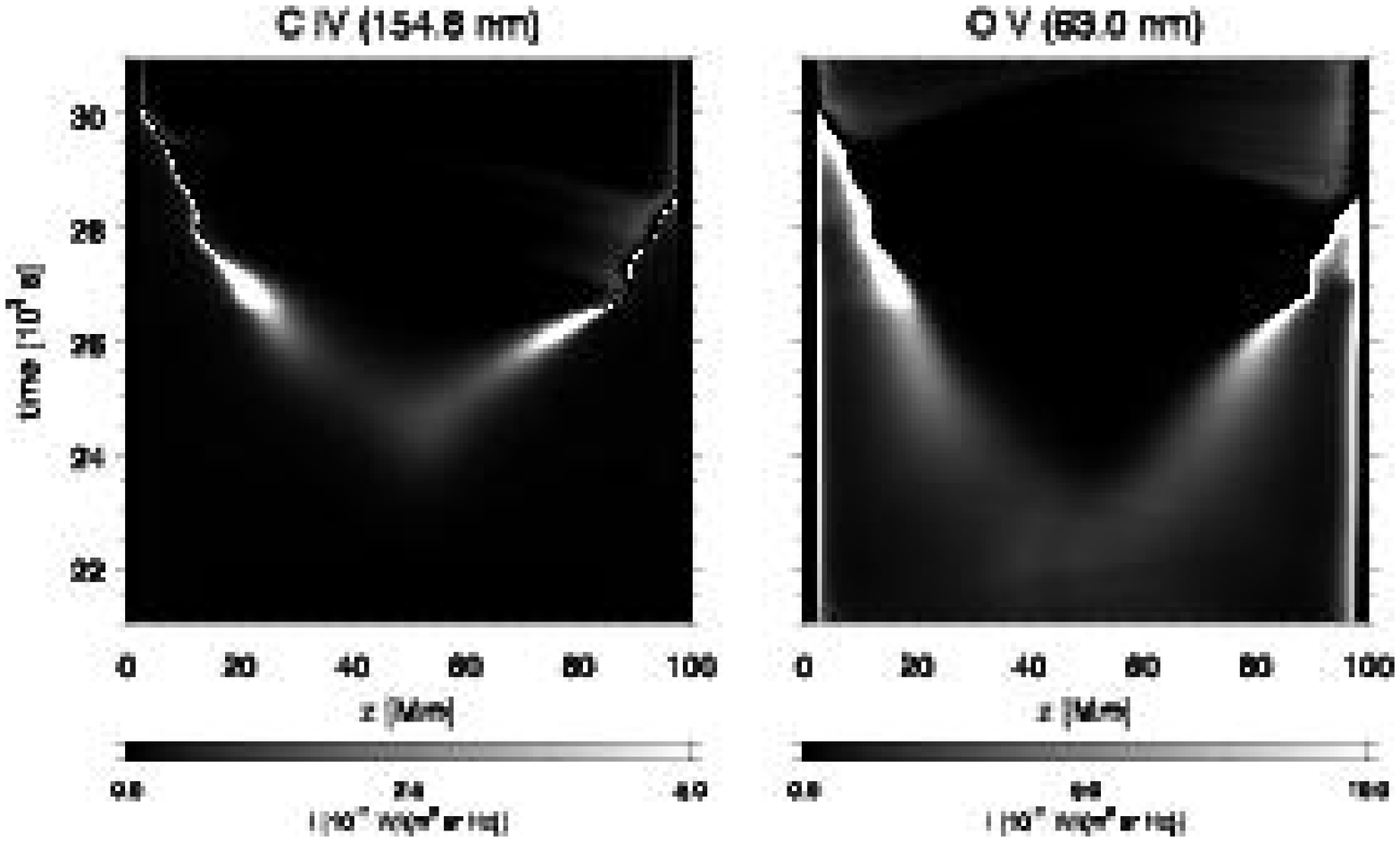}
\end{center}

\caption{\label{fig3a} Formation of two condensation regions in a coronal loop for $H_m = 2$\,Mm. The \emph{upper left} plot shows the evolution of temperature along the loop, the \emph{upper right} plot shows the corresponding velocities. The \emph{lower left} plot displays the emission in \ion{C}{4} (154.8\,nm), the \emph{lower right} plot the emission in \ion{O}{5} (63.0\,nm.)}
\end{figure*}

For an overview of temperature, velocity and emission for a small part of the simulation run with $H_m = 2$\,Mm, Fig.~\ref{fig3a} shows a cutout from Fig.~\ref{fig3} (\emph{right} panel), together with the velocity field and the corresponding emission in the two spectral lines of  \ion{C}{4} (154.8\,nm) and \ion{O}{5} (63.0\,nm). It is seen that the condensation regions are accompanied by strong transient brightenings in both lines. As  the \ion{O}{5} (63.0\,nm) line is formed at higher temperatures than the  \ion{C}{4} (154.8\,nm) line, a small time delay is observed between the occurrence of the brightenings in the two lines.
The wiggles in the path of the condensation region are due to the strong deceleration of the blob by the transition region plasma (cf.~Sect.~\ref{sec:vel_accel}).

\begin{figure}
\begin{center}
\includegraphics[width=8cm]{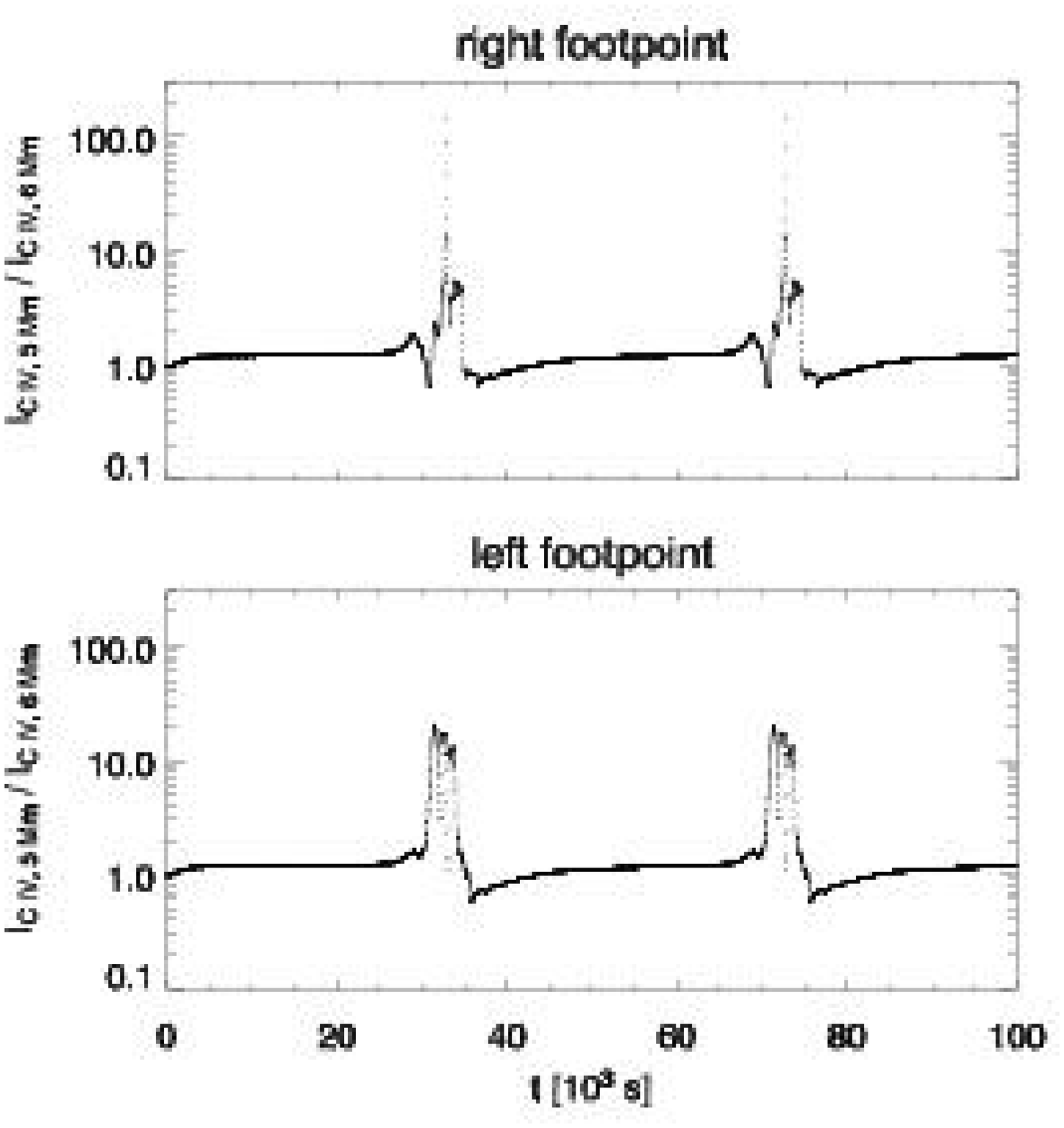}
\end{center}
\caption{\label{cfoot}Relative intensity, $I_{H_m = 5\,{\rm Mm}} / I_{H_m = 6\,{\rm Mm}}$, in \ion{C}{4} (154.8\,nm) at the left (\emph{upper} panel) and right (\emph{lower} panel) footpoints.}
\end{figure}

When the condensed plasma blob falls down the leg of the loop, it compresses the underlying plasma, which results in a transient temperature rise of the plasma underneath and a strong brightening around the footpoint of the loop when the plasma blob encounters the transition region.
Fig.~\ref{cfoot} shows the variation of the emission in \ion{C}{4} (154.8\,nm) at the loop footpoints ($z = 2$\,Mm and $z = 98$\,Mm) as a function of time. The intensity from this highly dynamic model run with $H_m=5$\,Mm is scaled by the respective intensity from the stable model run for $H_m=6$\,Mm (cf.\ Fig.~\ref{fig2}). This is done in order to highlight the dynamics in the \ion{C}{4} emission.
It is observed that the intensity at the right footpoint (draining direction of the condensation region, \emph{lower} panel) increases for a short time by more than two orders of magnitude and by more than one order of magnitude at the left footpoint. The latter effect takes place because the rarefaction wave following the falling condensation region pulls up plasma from the lower transition region to higher temperatures which leads to the strong transient brightening in the \ion{C}{4} (154.8\,nm) line.

\section{\label{sec:obs}Comparison to observations and discussion}
Several features of the numerical simulations presented here are in good agreement with recent observations from different instruments, so that we propose the condensation-evaporation cycle as a possible common explanation.

Let us briefly sum up the observational evidence and its analysis:
The TRACE observations of \cite{Schrijver2001SP} show strong transient brightenings in Ly$_\alpha$ and \ion{C}{4} (154.8\,nm), developing initially near the loop tops. Thereafter cool plasma slides down on both sides of the loops, forming clumps which move with velocities of up to 100\,km/s but show a downward acceleration of $80 \pm 30$\,m/s$^2$, significantly less than the solar surface gravity.
After a detailed analysis, \cite{Schrijver2001SP} concluded that the observed brightenings are due to the radiation of relatively dense blobs of falling plasma which are ``embedded in more tenuous cool matter or in plasma at a different temperature''. He argued that the reduced acceleration may be caused by the cooling plasma underneath the radiating blobs which could slow down the fall. Referring to the work of \cite{Mok+al1990ApJ}, he suggested that the observed catastrophic cooling could be explained by a drastic and fast reduction of the heating scale height which would result in a strong decrease of the heating at the loop apex.

\cite{DeGroof+al2004AA} analyzed a high cadence time series of simultaneous EIT (30.4\,nm) and Big Bear $H_\alpha$ data and found intensity variations in a coronal loop which propagated from the top towards the footpoint. The measured speeds of the blobs are compatible with a free-fall in the upper part of the loop but are significantly smaller in the lower part of the loop. Testing different hypotheses concerning the origin of the intensity variations, the authors rejected slow magneto-acoustic waves as an explanation for the observations. Instead, they favored flowing plasma blobs to account for the observed intensity variations.

Our simulations strongly support catastrophic cooling as the key mechanism to explain these sets of observations and provide further insight into the physical processes. In contrast to the work of \cite{Mok+al1990ApJ} we are able to synthesize optically thin emission lines forming in the transition region and corona, and can thus directly reproduce the transient brightenings in, e.g., \ion{C}{4} (154.8\,nm). We can also prove the suggestion of \cite{Schrijver2001SP} that the falling plasma blobs are decelerated by the underlying plasma and obtain quantitative agreement for the acceleration of the blobs. Furthermore, we find that this region is strongly compressed by the falling condensation region which leads to a strong transient brightening of the loop footpoint.
 
The main novelty that our simulations provide, however, is the finding that catastrophic cooling is not necessarily initiated by a sudden decrease of the heating or the heating scale height.
We support the presumption of \cite{Schrijver2001SP} that a ``drastic and fast reduction of the heating scale height suffices'' to trigger the formation of cool condensations. In fact, this is what is happening in the initial phase of all dynamic loop simulations presented here, when the heating scale height is instantaneously reduced.
Moreover, we show that catastrophic cooling does not have to be the result of a time-dependent heating scale height,
but can also result from a slowly evolving loss of equilibrium  at the loop apex as a natural consequence of loop heating predominantly at the footpoints.

On the other hand, it is hardly possible to trigger catastrophic cooling at all by just decreasing the amount of heating if this decrease is not accompanied by a decrease of the heating scale height (cf.~Sect.~\ref{sec:types}).

A small heating scale height rather than a heating function with time-dependent amplitude thus seems to be the key element for catastrophic cooling. 
These statements are important as they show that time-dependent phenomena observed in coronal loops do not demand time-dependent driving mechanisms (although many of them exist) but can also be the result of basic radiative or hydrodynamic instabilities.

The question has been raised whether a higher temperature of the loops would significantly alter their evolution. We have therefore carried out an additional set of simulations where we increased the mechanical energy flux to $F_{m0} = c \cdot 10^4$~W/m$^2$ while keeping all other parameters constant. This results in a start model with $T_{\rm max} = 2.7$\,MK and $\langle T \rangle = 2.5$\,MK and recurrently condensing loops with maximal temperatures of $\langle T \rangle_{\rm max} = 1.8 - 1.9$\,MK.
\begin{figure}
\resizebox{\hsize}{!}{
\includegraphics{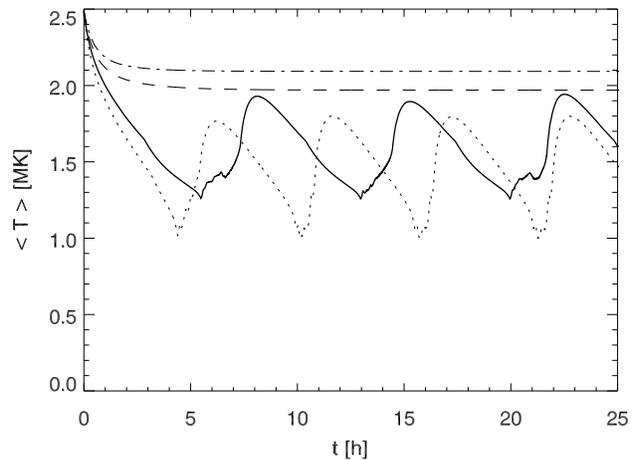}
}
\caption{\label{fig_mte_hot}Evolution of mean temperature, $\langle T \rangle (t)$, using an increased mechanical energy flux of $F_{m0} = c \cdot 10^4$~W/m$^2$, for four different damping lengths of the heating function: $H_m = 2$\,Mm (\emph{dotted}), $H_m = 3$\,Mm (\emph{solid}), $H_m = 5$\,Mm (\emph{dashed}), $H_m = 6$\,Mm (\emph{dash-dotted}).} 
\end{figure}
The evolution of the mean temperature of these loop models is displayed in Fig.~\ref{fig_mte_hot}. It is seen that the general behavior remains unchanged, in the sense that loops with a heating scale height below a certain threshold undergo periodic evolution. Compared to cooler loops with a lower heating rate, the increased heating rate results in a slight reduction of this threshold value, which is the expected result.
It has to be kept in mind that the loop length also significantly affects the maximal loop temperature so that longer loops reach much higher temperatures for a given mechanical energy flux.

Comparing the results of the current model with observations, we have to stress that the observed blob speeds are significantly smaller than the observed ones and the periods are lower than the time scale estimated by \cite{Schrijver2001SP}. However, as mentioned in Sect.~\ref{sec:vel_accel}, new simulations for a $L=300$\,Mm model loop (which corresponds to the estimated length of the loop analyzed by \cite{DeGroof+al2004AA}) yields blob speeds of the order of $100$\,km/s and periods of up to several days. A detailed comparison of these results with observational data is in progress \citep[][in preparation]{Mueller+al2004b}.

In this model we make the implicit assumption that the heating rate is not
affected by the catastrophic cooling, even though the density and gas
pressure change significantly while the condensation sets in (cf., e.g., the
average pressure in Fig.~\ref{fig4} or the density in Figs.~\ref{h3ana}
and \ref{h5ana}).
However, even during the condensation phase the plasma-$\beta$ remains
below 0.03 when assuming a reasonable value of
10\,Gauss for the magnetic field.
Thus, throughout we deal with a low-$\beta$ plasma, where the magnetic
field is presumably unperturbed by the plasma.

Therefore, when assuming a magnetically dominated heating mechanism like
flux-braiding \citep{Galsgaard+Nordlund1996JGR,Gudiksen+Nordlund2002ApJL} we expect the heating rate to remain constant (and on average decaying exponentially with height), regardless of the dynamic evolution of the plasma.

\section{Summary}

Our model calculations of coronal loops reproduce observations of catastrophic cooling and high-speed downflows, using a very simple, time-independent heating function. The non-linearity of the energy equation results in a loss of equilibrium which triggers a highly dynamic loop evolution. No external time-dependent driving mechanism is necessary to explain rapid cooling and evacuation of loops.
Coronal loops can exhibit cyclic behavior, with a wide range of periods, as well as irregular solutions. As our code solves the non-equilibrium rate equations consistently with the dynamic equations, the time-dependent emission of optically thin spectral lines can be synthesized and directly compared to observations giving a good match to the observed properties of catastrophic cooling of coronal loops.

\begin{acknowledgements} D.M. thanks the members of the Institute of
Theoretical Astrophysics, Oslo, for  their hospitality and support, and
acknowledges grants by the Deut\-sche For\-schungs\-ge\-mein\-schaft,
DFG, and the German National Merit Foundation. The authors thank the referee, Dr Karel Schrijver, for his detailed and helpful comments.
This work was also supported in part by the EU-Network HPRN-CT-2002-00310. 

\end{acknowledgements}

\bibliographystyle{aa}

\end{document}